\newcommand{\refalg}[1]{Algorithm~\ref{#1}}
\newcommand{\refsec}[1]{Sect.~\ref{#1}}
\newcommand{\reffig}[1]{Fig.~\ref{#1}}

\newcommand{\reftab}[1]{Tab.~\ref{#1}}
\newcommand{\refeqn}[1]{Eq.~(\ref{#1})}

\documentclass{sig-alternate}

\usepackage{subfigure}
\usepackage{enumitem}

\begin{document}


\CopyrightYear{2016}
\setcopyright{none}
\conferenceinfo{EASC '16,}{April 26 - 29, 2016, Stockholm, Sweden}

\title{On the Strong Scaling of the Spectral Element Solver Nek5000 on Petascale Systems}
%
%
%
%
%

\numberofauthors{10} 
%
\author{
%
%
\alignauthor
Nicolas Offermans\\
       \affaddr{Linn\'{e} FLOW Center}\\
       \affaddr{KTH Mechanics, Royal Institute of Technology}\\
       \affaddr{10044 Stockholm, Sweden}\\
       \email{nof@mech.kth.se}
\alignauthor
Oana Marin\\
       \affaddr{Mathematics and Computer Science Division}\\
       \affaddr{Argonne National Laboratory}\\
       \affaddr{Argonne, IL, USA}\\
       \email{oanam@mcs.anl.gov}
\alignauthor 
Michel Schanen\\
       \affaddr{Mathematics and Computer Science Division}\\
       \affaddr{Argonne National Laboratory}\\
       \affaddr{Argonne, IL, USA}\\
       \email{mschanen@anl.gov}
\and  
\alignauthor
Jing Gong\\
       \affaddr{PDC-HPC}\\
       \affaddr{KTH, Royal Institute of Technology}\\
       \affaddr{10044 Stockholm, Sweden}\\
       \email{gongjing@kth.se}
\alignauthor 
Paul Fischer\\
       \affaddr{Siebel Center for Computer Science}\\
       \affaddr{University of Illinois at Urbana-Champaign}\\
       \affaddr{Urbana, IL, USA}\\
       \email{fischerp@illinois.edu}
\alignauthor 
Philipp Schlatter\\
       \affaddr{Linn\'{e} FLOW Center}\\
       \affaddr{KTH Mechanics, Royal Institute of Technology}\\
       \affaddr{10044 Stockholm, Sweden}\\
       \email{pschlatt@mech.kth.se}
}
\additionalauthors{
Aleks Obabko (ANL, email: {\texttt{obabko@mcs.anl.gov}}), \\
Adam Peplinski (KTH, email: {\texttt{adam@mech.kth.se}}), \\
Maxwell Hutchinson (University of Chicago, email:  {\texttt{maxhutch@uchicago.edu}}) and\\
Elia Merzari (ANL, email: {\texttt{emerzari@anl.gov}}).}
\date{21 March 2016}

\maketitle
\begin{abstract}
The present work is targeted at performing a strong scaling study of the high-order 
spectral element fluid dynamics solver Nek5000.  Prior studies such as 
\cite{fischer:scaling} indicated a recommendable metric for strong scalability 
from a theoretical viewpoint, which we test here extensively on three parallel 
machines with different performance characteristics and interconnect networks, 
namely Mira (IBM Blue Gene/Q), Beskow (Cray XC40) and Titan (Cray XK7). The test 
cases considered for the simulations correspond to a turbulent flow in a straight 
pipe at four different friction Reynolds numbers $Re_{\tau} = 180$, $360$, $550$ 
and $1000$.  Considering the linear model for parallel communication we quantify 
the machine characteristics in order to better assess the scaling behaviors of 
the code. Subsequently
sampling and profiling tools are used to measure the computation and communication 
times over a large range of compute cores.
We also study the effect of the two coarse grid solvers XXT and AMG on the
computational time. Super-linear scaling due to a reduction in cache misses is
observed on each computer. The strong scaling limit is attained for roughly
$5000-10,000$ degrees of freedom per core on Mira, $30,000-50,0000$ on Beskow,
with only a small impact of the problem size for both machines, and ranges
between $10,000$ and $220,000$ depending on the problem size on Titan. This work
aims at being a reference for Nek5000 users and also serves as a basis for
potential issues to address as the community heads towards exascale supercomputers.
\end{abstract}

\keywords{Computational Fluid Dynamics; Nek5000; Scaling; Benchmarking}

\section{Introduction}
The development of highly scalable codes that perform well on different
architectures has been a daunting task ever since the advent of high performance
computing, due to the interplay between computation and communication,
inescapable global operations but foremost due to the nature of this research
field constantly redefining its path. In the current work we explore the parallelism 
of Nek5000, which is one of the oldest legacy codes (celebrating 30 years this year) 
and thus has experienced many trends and changes in high performance computing strategies.

Nek5000 is a code based on the spectral element method, intended to solve problems
from thermal hydraulics, which performs best on
complex geometries, wall-bounded problems (although it can handle the most common types of
boundary conditions), at large scales on any commonly used parallel computer
architecture. The present study is aimed at providing users a handle on
parameter choices for performance and scalability, and relies on previous work ,
such as \cite{fischer:scaling} and \cite{tufo:terascale}. Hereby we benchmark
the code on a canonical flow case, a direct numerical simulation (DNS) of
the incompressible flow in a pipe at increasingly high Reynolds numbers
\cite{Khoury2013}. Solving a Poisson-like equation for the
pressure is commonly the most challenging computational part of an incompressible flow
solver. Nek5000 relies on the construction of an efficient preconditioner to solving the Poisson
subproblem. This preconditioner is obtained by combining a domain decomposition 
approach and a coarse grid solve being computed either via XXT \cite{Tufo2001151} 
or AMG \cite{LottesAMG}. We explore both 
approaches and quantify the regimes in which either of them is recommendable. 

In the rest of the paper, we start by giving a short description of the numerical method
and implementation. Then we describe the hardware employed, focusing particularly on
the architecture, interconnect network technology and associated latency and
bandwidth. We also present the performance analysis tools we used for profiling the code
as well as the test cases considered for performing the tests. We finish with a 
description of results, we identify the strong scaling limit, discuss about the
observed super-linearity and compare the two coarse grid solvers XXT and AMG. For 
a more complete interpretation of the results we assess also load balancing, 
mesh partitioning, cache misses etc.

\section{Code description}
Nek5000 supports a wide set of options that speed up the time to solution, such
as the method of characteristics which decouples the pressure solve from the
restrictive CFL condition for the nonlinear advection operator, or orthogonal
projections of the solution to reduce the iteration count of the algebraic
solver etc. Here we focus only on one track to solution which is consistent with
the physical case we study and the way it was initially performed, i.e. fully 
resolved DNS of a turbulent pipe flow \cite{Khoury2013}.

\subsection{Numerical method}
\label{sec:method}
The incompressible Navier--Stokes equations are given here by
\begin{align} 
 \frac{\partial \mathbf{u}}{\partial t} + (\mathbf{u \cdot \nabla}) \mathbf{u} & = - \nabla p + \frac{1}{Re} \nabla^2 \mathbf{u} + \mathbf{f} \label{eqn:NS_momentum},\\
 \nabla \cdot \mathbf{u} & = 0, \label{eqn:NS_continuity}
\end{align}
where $\mathbf{u}$ is the velocity, $p$ the pressure and $\mathbf{f}$ a forcing 
term. The Reynolds number 
$Re = \frac{U L}{\nu}$ is expressed as a function of a typical velocity scale $U$,
length scale $L$ and kinematic viscosity $\nu$. \refeqn{eqn:NS_momentum} 
and \refeqn{eqn:NS_continuity} are called the continuity and momentum equations 
respectively. There are two main solvers, called PNPN ($\mathbb{P}_N-\mathbb{P}_N$) and PNPN-2 
($\mathbb{P}_N-\mathbb{P}_{N-2}$), available within 
Nek5000 for computing the solution of the incompressible 
Navier-Stokes equations, and of these one is also amenable to non-divergence free 
flows, as available in \cite{Tomboulides1997}, namely $\mathbb{P}_N-\mathbb{P}_N$. Although we operate in the incompressible
regime we picked this solver to preserve generality. 

The momentum equation is time integrated via an implicit-explicit scheme, also
known as BDFk-EXTk (Backward Difference formula and EXTrapolation of order k). We
illustrate it semi-discretely as

\begin{eqnarray}
\sum\limits_{j=0}^k \frac{b_j}{\Delta t} \mathbf u^{n-j}  = - \nabla p^{n}+\frac{1}{Re}\nabla^2\mathbf u^{n}+\underbrace{\sum\limits_{j=1}^k a_j [N(\mathbf u^{n-j})+\mathbf f^{n}]}_{\mathbf{F}_n(\mathbf u,\mathbf f)},\label{eqn:discrete}
\end{eqnarray}
where we denoted the nonlinear operator $\mathbf u \cdot \nabla \mathbf u=N(\mathbf u)$ and $b_k$, $a_k$ are the coefficients of the implicit time derivative discretization, and explicit extrapolation respectively.

Ignoring boundary conditions and other numerical technicalities available in \cite{Tomboulides1997} we end up solving
\begin{eqnarray}
 \Delta p^{n} &= &\nabla \cdot  \mathbf{F}_n \left( \mathbf{u},\mathbf f \right), \label{eqn:hmhz_pres}\\
 \frac{1}{Re}\Delta \mathbf{u}^{n}- \frac{b_0}{\Delta t} \mathbf{u}^{n} & = & \nabla p^{n} + \mathbf{F}_n ( \mathbf{u}, \mathbf f)  +\sum\limits_{j=1}^k \frac{b_j}{\Delta t} \mathbf u^{n-j} . \label{eqn:hmhz_vel}
\end{eqnarray}

\begin{figure}
  \centering
  \subfigure[Partition]{
  \includegraphics[trim=300 420 800 360,clip,width=\linewidth]{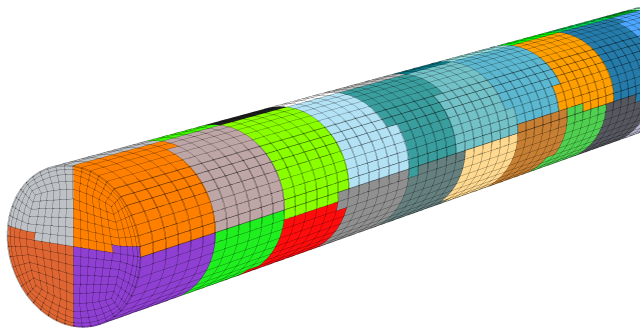}
  \label{fig:partition_vis}
  }
  \subfigure[Velocity magnitude]{
  \includegraphics[trim=300 420 800 360,clip,width=\linewidth]{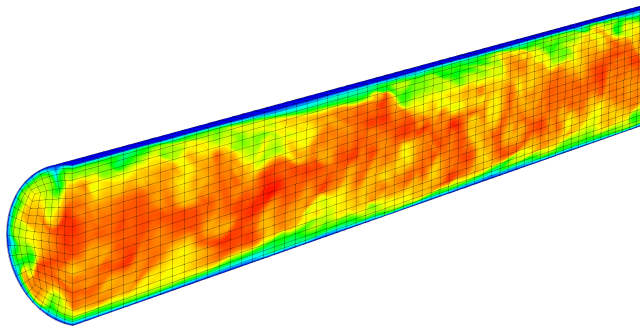}
  \label{fig:flow_vis}
  }
  \caption{Partition of the elements on $64$ processes and velocity magnitude in the pipe ($Re_{\tau}=180$).}
  \label{fig:partition}
\end{figure}

As it can be observed, solving the incompressible Navier-Stokes equations is reduced 
to the evaluation of $\mathbf{F}_n$ in \refeqn{eqn:discrete}, followed by one Poisson 
equation and a Helmholtz equation thereafter for each velocity component (2 in 2D and 3 in 3D). 
\refeqn{eqn:hmhz_pres}, the Poisson equation for the pressure, is the main 
source of stiffness and its efficient resolution by an iterative solver is preceded 
by two steps. First of all, the pressure at each time step is projected onto a 
subspace of previous solutions, and as described in \cite{Fischer1998} has been shown 
to reduce the iteration count by a factor $2.5-5$, which we also verify in \refsec{sec:analysis}.
Then, a pressure preconditioner is built based on the additive overlapping Schwarz 
method, given by 
\begin{equation}
 M_0^{-1} := R_0^T A_{0}^{-1} R_0 + \sum_{k=1}^{K} R_k^T \tilde{A}_k^{-1} R_k.
\end{equation}
The overlapping part requires local solves on each subdomain and is naturally parallelizable 
despite a fairly complex practical implementation \cite{Fischer199784,Fischer2005}. 
The coarse grid solve is in essence more difficult to parallelize and this can be 
performed in two different ways. The first 
method is a Cholesky factorization of the matrix $A_0^{-1}$ into $XX^T$ with a 
convenient refactoring of the underlying matrix to maximize the sparsity pattern 
of $X^T$. This factorization is subsequently referred to as XXT and details 
regarding complexity and implementation are available in \cite{Tufo2001151}. The 
second method is a single V-cycle of a highly-tuned
AMG solver that is designed specifically to be communication minimal
and optimal for coarse-grid problems where one anticipates very
few degrees of freedom per processor \cite{LottesAMG}.

\subsection{Implementation}
\label{sec:implementation}

\subsubsection{Mesh and mapping}
\label{sec:mesh_map}

The geometry is meshed using hexahedral elements, partitioned for parallel
computation using a spectral bisection algorithm as implemented in ``genmap''
which accompanies the code Nek5000 \cite{argonne:nekdoc}. An example of the
partitioning for the case $Re_{\tau} = 180$ run on $64$ cores is shown in \reffig{fig:partition_vis}, where each element is colored according to the MPI rank it belongs to. We note that the partitioning is done at the element level and not finer.
 
\subsubsection{Code structure}
\label{sec:code}

The sequence of operations leading to the solution of the incompressible
Navier--Stokes equations is summed up algorithmically in \refalg{alg:code_struct}.  
First of all, note that both \refeqn{eqn:hmhz_pres} and \refeqn{eqn:hmhz_vel} can be summed up in discrete form as
$$H \phi=(h_1 A+h_2 B)[\phi],$$
where $A$ is the stiffness matrix stemming from the discretization of the Laplacian and $B$ is the mass matrix. 
Different choices for the factors $h_1$ and $h_2$ yield either the Poisson equation, or the Helmholtz equation
\begin{eqnarray}
Hp &=& (A)[p]=f_p, \quad (h_1=1, h_2=0)\\
Hu &=&(\frac{1}{Re}A -\frac{b_0}{\Delta t} B )[u]=f_u, \quad (h_1=1, h_2=-\frac{b_0}{\Delta t}).
\end{eqnarray}

\begin{algorithm}
\begin{algorithmic}
\Procedure{Solver}{}
\For {k=1,...,nsteps} 
\State \# \textbf{Compute Pressure}
\State $f_p \leftarrow rhs_p(u, \mathbf{f})$ 
\State $\delta f_p \leftarrow f_p - \text{proj}_{X_{f_p}}(f_p)$ 
\State $\delta p \leftarrow \text{Helmholtz}(H_p,\delta f_p)$ 
\State $p \leftarrow p + \delta p$ 
\State $X_p \leftarrow \left\{ X_p, \text{proj}_{X_{p}}(p) \right\}$ 
\State $X_{f_p} \leftarrow \left\{ X_{f_p}, \text{proj}_{X_{f_p}}(H_p p) \right\}$ 
\State \# \textbf{Compute Velocity}
\State $f_{u} \leftarrow rhs_{u}(p, u, \mathbf{f})$ 
\State $\delta f_{u} \leftarrow f_{u} - \text{proj}_{X_{f_{u}}}(f_{u})$ 
\State $\delta {u} \leftarrow \text{Helmholtz}(H_{u},\delta f_{u})$ 
\State ${u} \leftarrow u + \delta u$ 
\State $X_{u} \leftarrow \left\{ X_{u}, \text{proj}_{X_{u}}(u) \right\}$ 
\State $X_{r_{u}} \leftarrow \left\{ X_{f_{u}}, \text{proj}_{X_{f_u}}(H_{u} u) \right\}$ 
\EndFor
\EndProcedure
\end{algorithmic}
\caption{Main solver.}
\label{alg:code_struct}
\end{algorithm}

By virtue of the method of projections, we do not solve $Hp=f_p$ or $Hu=f_u$, but 
rather $H\delta p=\delta f_p$ and $H\delta u=\delta f_u$, where $\delta f_p$ and 
$\delta f_u$ are the rejections of $f_p$ and $f_u$ respectively. Details on the 
technicalities of this are abundant in \cite{Fischer1998}. The first step is to 
compute the corresponding right hand sides and then project them onto a subspace 
of previous solutions (subspaces are denoted by $X$ and the size of the space is $L$).
 The corrections $\delta p$ and $\delta u$ are then computed by solving the Helmholtz 
equation for the rejections and added to the previous solution. A simplified structure 
for the Helmholtz solver is shown in \refalg{alg:helmholtz}. The Poisson equation for
the pressure is solved with the GMRES method. The pressure solve also includes the
computation of the preconditioner based on the Schwarz overlapping method and 
coarse grid solve, which is not the case for the velocity and constitutes an 
important part of the work and communication. The Helmholtz equation is solved using the CG method for each component of the velocity.

\begin{algorithm}
\begin{algorithmic}
\Procedure{Helmholtz}{$H, r$}
\If {Velocity}
\State $x \leftarrow CG(H, r)$
\ElsIf {Pressure}
\State $\left(M_0^{-1}\right)_{\text{Sch}} \leftarrow$ Overlapping Schwarz()
\State $\left(M_0^{-1}\right)_{\text{cgs}} \leftarrow$ Coarse grid solve() 
\State $M_0^{-1} \leftarrow \left(M_0^{-1}\right)_{\text{Sch}} + \left(M_0^{-1}\right)_{\text{cgs}}$
\State $x \leftarrow GMRES(M_0^{-1}, H, r)$
\EndIf
\State \textbf{return} x
\EndProcedure
\end{algorithmic}
\caption{Helmholtz solver.}
\label{alg:helmholtz}
\end{algorithm}

\section{Benchmarking}
\label{sec:benchmarking}

\subsection{Hardware}
\label{sec:hardware}

The test cases were run on three different supercomputers, namely Mira from the
Argonne National Laboratory, USA, Titan from the Oak Ridge National Laboratory,
USA, and Beskow from the PDC Center for High Performance Computing, KTH, Sweden.
A quick overview of the characteristics of each computer is summarized in 
\reftab{tab:computer_charac}.
\begin{table*}
\centering
\caption{Overview of the characteristics of the different supercomputers.}
\begin{tabular}{l|cccccc} 
\hline
 & System arch. & Core arch. & Number of cores & Cores/node & Topology & Processes/core \\
 \hline
Mira   & IBM BG/Q & PowerPC A2 & $786,432$ & $16$ & 5D torus  &2\\ 
Titan  & Cray XK7 & AMD Opteron & $299,008$ & $16$ & 3D torus  &1\\ 
Beskow & Cray XC40 & Intel Haswell & $53,632$  & $32$ & DragonFly &1\\
\hline
\end{tabular}
\label{tab:computer_charac}
\end{table*}
The systems vary from small to large petascale and are meant to establish an overview of the Nek5000 scaling. On Mira, Nek5000 achieves its
maximum performance when run with two processes per BG/Q core, being 32 processes
per node which was noted already in \cite{fischer:scaling}. Although Titan is a
machine aimed at hybrid parallelism using graphics processing units (GPUs), which
Nek5000 supports marginally as mentioned in \cite{Otten2016}, no production runs
were performed outside the MPI environment and we shall restrict the use of
Titan to CPU parallelism and rely solely
on the 16 Opteron cores per node with one process each. The same setup of 1
process per CPU core, i.e. not using hyperthreading, was applied to the smallest 
system Beskow, a Cray XC40. Indeed, some tests performed on a single Haswell core 
showed that hyperthreading did not improve time to solution.

\begin{table}
\centering
\caption{Overview of the latencies and bandwidths. A word (wd) is 64 bits long.}
\begin{tabular}{l|ccccc} 
\hline
 & $\alpha^* (\unit[]{\mu s})$ & $\beta^* (\unit[]{\mu s/wd})$ & $t_a (\unit[]{\mu s})$ & $\alpha$ & $\beta$ \\
 \hline
 Mira   &  $4$   & $5\cdot 10^{-3}$                     & $1.1 \cdot 10^{-3}$ & $3600$ & $5$   \\ 
Titan  & $2.25$ & $1.42\cdot 10^{-3}$ & $6.5 \cdot 10^{-4}$ & $3500$ & $2.2$ \\
Beskow & $2.55$ & $8.25\cdot 10^{-4}$ & $1.5 \cdot 10^{-4}$ &$17000$ & $5.5$ \\ 
\hline
\end{tabular}
\label{tab:alpha_beta}
\end{table}

\begin{figure}
  \centering
  \subfigure[Ping-pong and all-reduce on Mira as illustrated in \cite{fischer:scaling}]{
  \includegraphics[trim=40 0 40 27,clip,width=\linewidth]{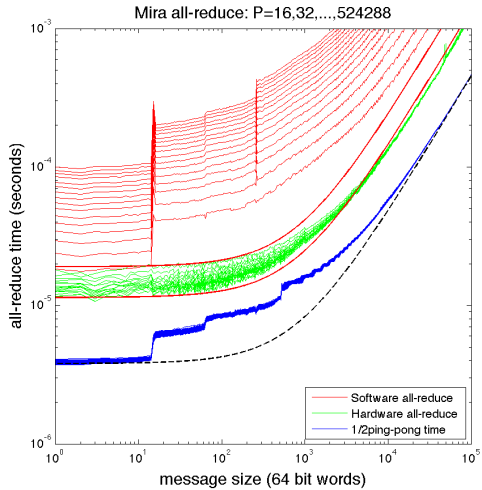}
  \label{fig:pingpong_mira}
  }
  \subfigure[Ping-pong on Titan]{
  \includegraphics[width=\linewidth]{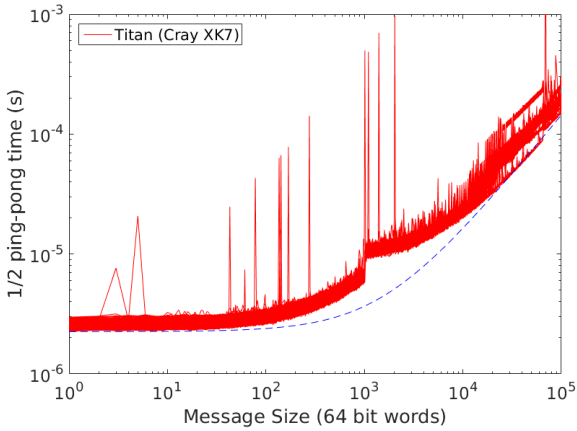}
  \label{fig:pingpong_titan}
  }
  \subfigure[Ping-pong on Beskow]{
  \includegraphics[width=\linewidth]{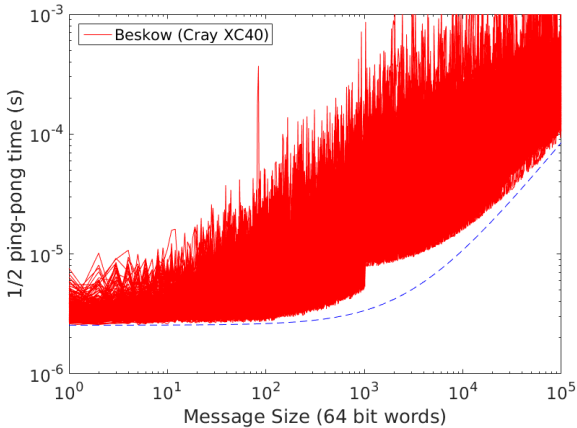}
  \label{fig:pingpong_beskow}
  }
  \caption{Latency and bandwidth tests.}
  \label{fig:pingpong}
\end{figure}

In order to assess the performance of the machines at hand, we computed  some of the network characteristics that determine the communication. In particular Beskow, which is a relatively new machine, had no such parameters provided to users. 
The performance study conducted here relies on the linear interprocessor communication  model
\begin{equation}
 t_c(m) = (\alpha + \beta m) t_a,\label{eqn:lincomm}
\end{equation}
where $t_c$ is the communication time, $m$ is the message length (number of
64-bit words) and $t_a$ is the inverse of the observed flop rate. The relevant
quantities here are $\alpha$ and $\beta$, the non-dimensional latency and inverse
 bandwidth. We denote by $\alpha^*$ and $\beta^*$ the corresponding dimensional 
latency and inverse bandwidth. The relation between dimensional and non-dimensional parameters is given by
\begin{equation*}
\alpha = \frac{\alpha^*}{t_a} \qquad \text{and} \qquad \beta = \frac{\beta^*}{t_a}.
\end{equation*}
The values of $\alpha^*$ and $\beta^*$ are computed following a ``ping-pong" test 
as described in \cite{fischer:scaling}. During this test, the time required to 
send and receive messages of various sizes between 512 processors (default value
for the test in Nek5000) is measured 
and subsequently the values of $\alpha$ and $\beta$ are computed as the best fit 
for the linear model \refeqn{eqn:lincomm}. The value of $t_a$ is determined by 
performing a number of matrix-matrix multiplications streamed from memory
representing the tensor products that are at the core of a spectral element 
solver \cite{fischer:hom} and accounting for a big part of the solve time 
\cite{Max2016}. The tensor products considered imply 3D elements with polynomial 
order ranging from $10$ to $13$. Three different interpretations of the memory 
layout for the matrices are considered 
leading to a total of $12$ tests. For each test, the time and number of operations are
measured and flop rate is computed. Data are then averaged and $t_a$ is taken as the
inverse of the mean flop rate. Results for the ping-pong test are shown in 
\reffig{fig:pingpong} along with the linear model for all computers. An overview 
of the latencies, bandwidths and inverse flop rates is presented in 
\reftab{tab:alpha_beta}. The non-dimensional parameters $\alpha$ and $\beta$ are 
a relative measure of the communication to computation cost. High values for these
parameters imply that the limitation in parallel efficiency will arise earlier due 
to a relatively high communication cost. Consistent throughout all our studies is 
that all machines are strongly limited by the latency, already a noted common 
feature of modern computers. This limitation is strongest on
Beskow, which is the newest of the three machines and has fastest CPUs. Therefore,
it is expected that Beskow will not scale as well as the two others.

A drawback of our performance model is that it does not capture system
noise. Furthermore, it assumes that all communication between two processes is
homogeneous; it does not distinguish for example between on node and off node
communication. This model works well for Mira. This work will also point out its
weakness when dealing with system noise. For a better understanding
of the impact of system noise, a relevant discussion about noise at 
scale can be found in \cite{Hoefler:2010}.

As a side note, hardware might not be the only responsible for the high noise in
communication and we would like to mention as an indication that we used the Cray 
programming environment version $5.2.26$.

\subsection{Code instrumentation for profiling}
\label{sec:timers}

We assess the scalability and parallel efficiency of Nek5000 by studying the distribution
between the time spent in communication and the time spent in computation for each 
simulation. These measurements are performed with performance tools adapted to each 
computer. The tools are set to start counting after the initial setup stage is 
completed, in our case after timestep 30, lasting for an extra 20 timesteps.  
The initial stage is meant to allow for the high-order restart, proper initialization 
of the projection space (i.e.  the size of the projection space $X$ is $L= 5$, thus
requiring 5 consecutive solutions). In order to measure the time spent in communication 
we relied on Craypat for
Beskow and Titan and on Hardware Performance Monitor (HPM) for Mira. Both tools
allow us to measure the total time spent in communication during the targeted 20 time
steps. HPM gives additional information on the cache misses and the load
imbalance. The CrayPat performance analysis framework is used to sample the code 
during execution at a default frequency of $\unit[100]{Hz}$ and reports in which 
function each sample was taken. Then, we assume that the proportion of the total 
time spent in a given function is equal to the proportion of samples within this 
function. The sampling procedure ensures a very low overhead. We also tested the 
tracing procedure, where all function calls are traced, available with Craypat 
but overhead in time was about $50\%$ and the method was abandoned.

\subsection{Test case : pipe flow}
\label{sec:pipe}

The test case considered is the turbulent flow in a straight pipe. A thorough 
description of the flow configuration as well as a detailed analysis of the physical 
results can be found in \cite{Khoury2013}. The flow was run at four different 
friction Reynolds numbers $Re_{\tau} = 180$, $360$, $550$ and $1000$. A summary 
of the different simulations and associated number of 
elements, polynomial order and number of grid points $N$ is presented in \reftab{tab:pipe_conf}. 
The friction Reynolds number is defined as $Re_{\tau} = u_{\tau} R / \nu$, where $u_{\tau}$ is 
the friction velocity, $R$ is the radius of the pipe and $\nu$ is the kinematic viscosity. The 
bulk Reynolds number is defined as $Re_{b} = 2 U_b R / \nu$, where $U_b$ is the mean bulk
velocity. A pressure gradient is imposed inside the pipe through a forcing term and periodic 
boundary conditions are imposed at the inlet and the outlet. A snapshot of 
the velocity magnitude for the case $Re_{\tau} = 180$ is illustrated in \reffig{fig:flow_vis}.

\begin{table}
\centering
\caption{Summary of the different pipe flows configurations.}
\begin{tabular}{llrcr} 
\hline
$Re_{\tau}$&$Re_{b}$&\# elements & pol. order & \# grid points\\ 
\hline
$180$ & $5300$ & $36,480$ & $8$ & $18.67 \times 10^6$\\
$360$ & $11,700$ & $237,120$ & $8$ & $121.4 \times 10^6$\\ 
$550$ & $19,000$ & $823,632$ & $8$ & $437.0 \times 10^6$\\ 
$1000$ & $37,700$ & $1,264,032$ & $12$ & $2.184 \times 10^9$\\
\hline
\end{tabular}
\label{tab:pipe_conf}
\end{table}

\section{Performance and Scaling Analysis}
\label{sec:analysis}
Our abstraction assumes that large scale runtime performance is mainly composed
of
\begin{enumerate}
  \item system hardware parameters consisting of the network topology, latency and
    bandwidth\label{enum:a}, and flops per second of matrix matrix products
    (usually memory bandwidth bound), 
  \item time $T_a$ and $T_c$ spent in computation and communication for the measured 
 $20$ timesteps largely dependent on point \ref{enum:a} and on their respective algorithmic complexities,
  \item partitioning.\label{enum:c}
\end{enumerate}

The partitioning for our test case (see \refsec{sec:pipe}) is considered to be
topologically equivalent to a cube. The resulting runtime complexities are
extensively described in \cite{fischer:scaling} for the Mira system. Based on
these theoretical results we use profiling tools and wall clock timers to
measure the load imbalance,
cache misses, as well as weak and strong scaling. Load imbalance and cache
misses are only measured on Mira via the HPM profiling library. 
We want in particular to verify
experimentally the strong scaling limit for a given problem of size $N$. 
In this paper, the strong scaling limit for a problem of size $N$ is defined
as the minimal number of grid points per process $\frac{N}{P}$ by finding $P$,
the number of processes, such that
\begin{equation}
  \dfrac{T_a(N,P)}{T_c(N,P)}=1,\mbox{ for problem size }N.
  \label{eq:strong}
\end{equation} 
\noindent Alternatively, the strong scaling limit is commonly described by the
derivative $\frac{dT}{dP}$
of the total time $T(N,P)$ being equal to $0$, i.e.\ the point where the runtime
starts increasing again with increasing $P$. That point represents the fastest
time to solution. This is rather an upper bound of the strong scaling
limit that would in most of our test runs never be observed. For Nek5000 on Mira it has
little practical meaning to the user, as that limit always implies the usage
of all of Mira, which is in most cases too costly. Without a cost model,
 \refeqn{eq:strong} gives us a much lower and practical bound of the strong
scaling limit. If communication takes longer than
computation, the code is deemed to be at the strong scaling limit where the
scaling starts diverging significantly from the perfect linear scaling. 

The test case used to explore the scaling behavior of Nek5000 is the one of a turbulent 
flow in a straight pipe, a generic and widely known case across the CFD community.
This should allow potential users to estimate and compare the scaling of Nek5000 to other CFD software.
Our test case is run in four different regimes for 4 different problem sizes denoted by
$Re_{\tau} = 180$, $360$, $550$ and $1000$, described more in
detail in \refsec{sec:pipe}. These cases were run with various processor counts 
on three systems. The lower bound of the processor count is dictated by the size 
of the random access memory (RAM) of each machine, i.e. the smallest number of 
nodes on which the problem can be packed. Nek5000 has roughly a memory requirement of 500 fields 
times the number of degrees of freedom. The upper bound was either set by the
administrative limit of getting access to the maximum number of processors
(Beskow and Titan) or by the algorithmic limit of having at least one element 
per process, since as noted in \refsec{sec:code}, the smallest parallelizable 
unit in Nek5000 is one spectral element. Of the 20 timesteps along which the
statistics are taken we consider the mean communication time reported over all
processes. 

In \cite{fischer:scaling} the strong scaling limit \refeqn{eq:strong} was estimated
to be around $\frac{N}{P}=2000$ for the conjugate gradients (CG) and
$7000$ for the geometric multigrid (MG) on Mira when using finite
differences. These values are given as an indication
but we remind that Nek5000 is an incompressible flow solver and gathers several different algebraic 
solvers each with its own customized preconditioning strategy. Indeed CG is implemented 
with Jacobi preconditioning for velocity and GMRES is used with XXT and AMG (which 
is different from the geometric multigrid) preconditioning for pressure. 
Taking also the added computational effort from projections, right hand side 
evaluations and the heavy communication of the coarse grid solver our values deviate slightly from the theoretical results in \cite{fischer:scaling}. 
\section{Results}

\begin{figure}
  \centering
  \subfigure[$Re_{\tau} = 180$]{
  \includegraphics[width=0.85\linewidth,height=0.2\textheight]{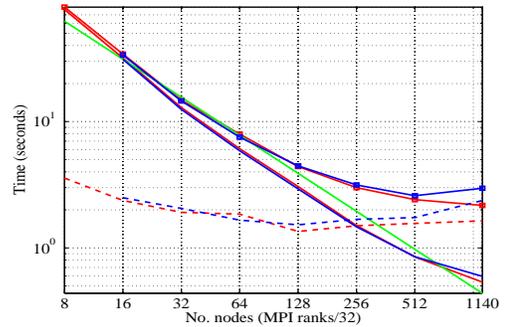}
  }
  \subfigure[$Re_{\tau} = 360$]{
  \includegraphics[width=0.85\linewidth,height=0.2\textheight]{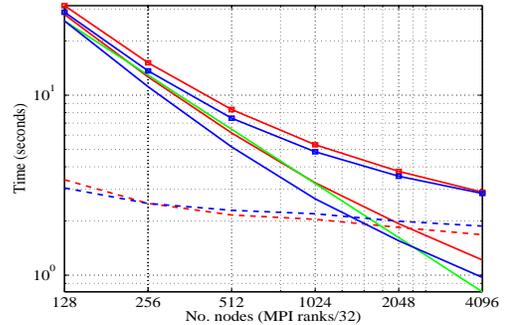}
  }
  \subfigure[$Re_{\tau} = 550$]{
  \includegraphics[width=0.85\linewidth,height=0.2\textheight]{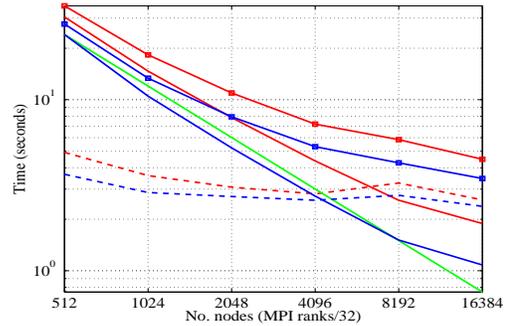}
  }
  \subfigure[$Re_{\tau} = 1000$]{
  \includegraphics[width=0.85\linewidth,height=0.2\textheight]{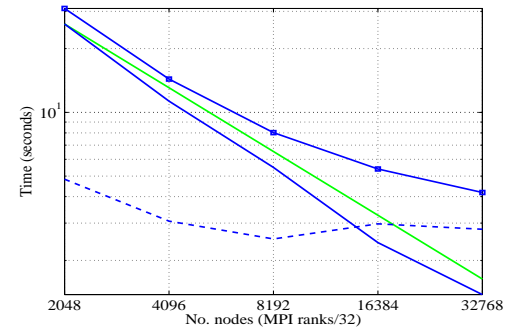}
  }
  \caption{Total time for 20 timesteps on BG/Q Mira: AMG ({\color{blue}blue}), XXT ({\color{red}red}), communication (dashed),
  computation (solid), total time ($\square$), computational linear scaling
  ({\color{green}green}).}
  \label{fig:scaling_mira}
\end{figure}

\begin{figure}
  \centering
  \subfigure[$Re_{\tau} = 180$]{
  \includegraphics[width=0.85\linewidth,height=0.2\textheight]{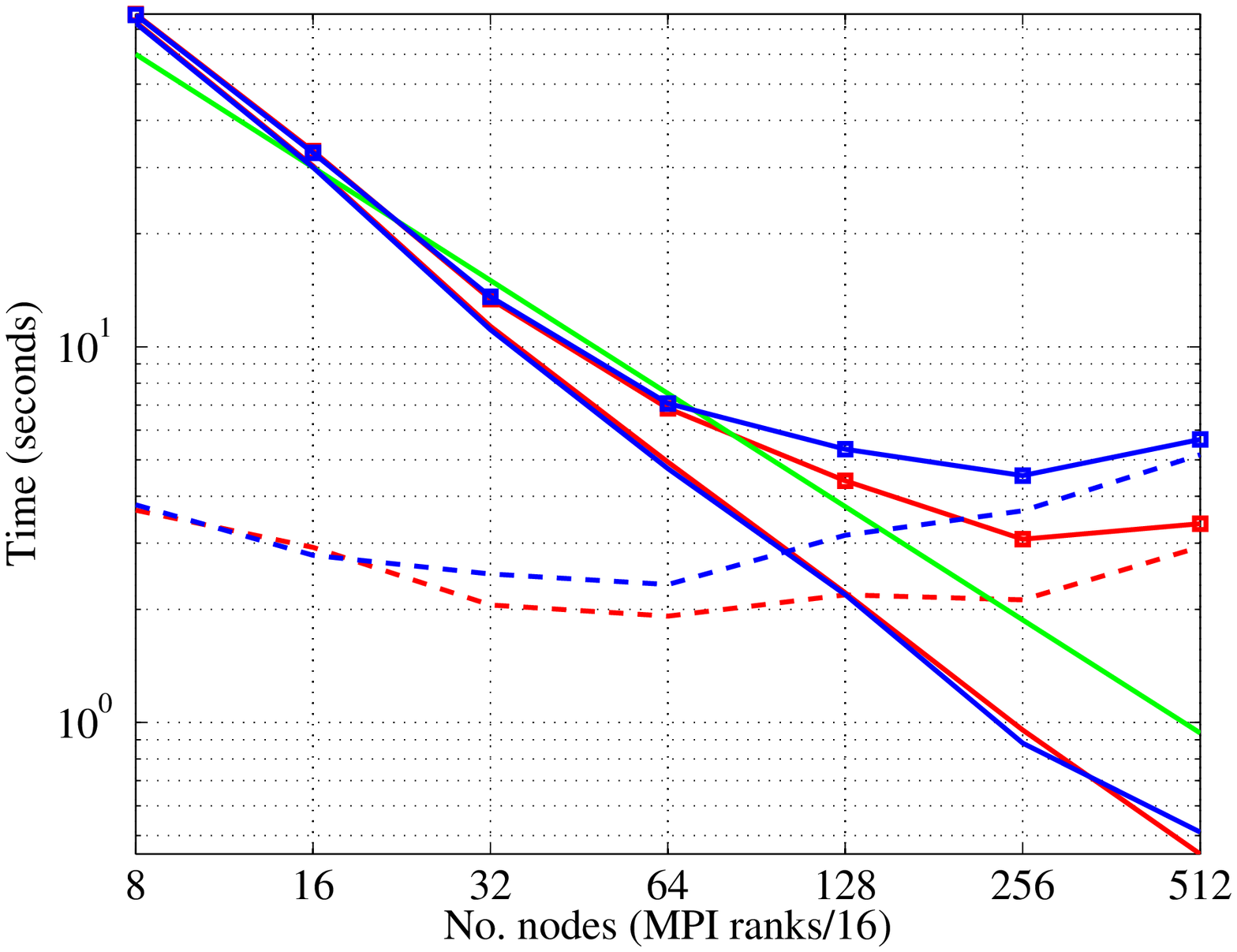}
  }
  \subfigure[$Re_{\tau} = 360$]{
  \includegraphics[width=0.85\linewidth,height=0.2\textheight]{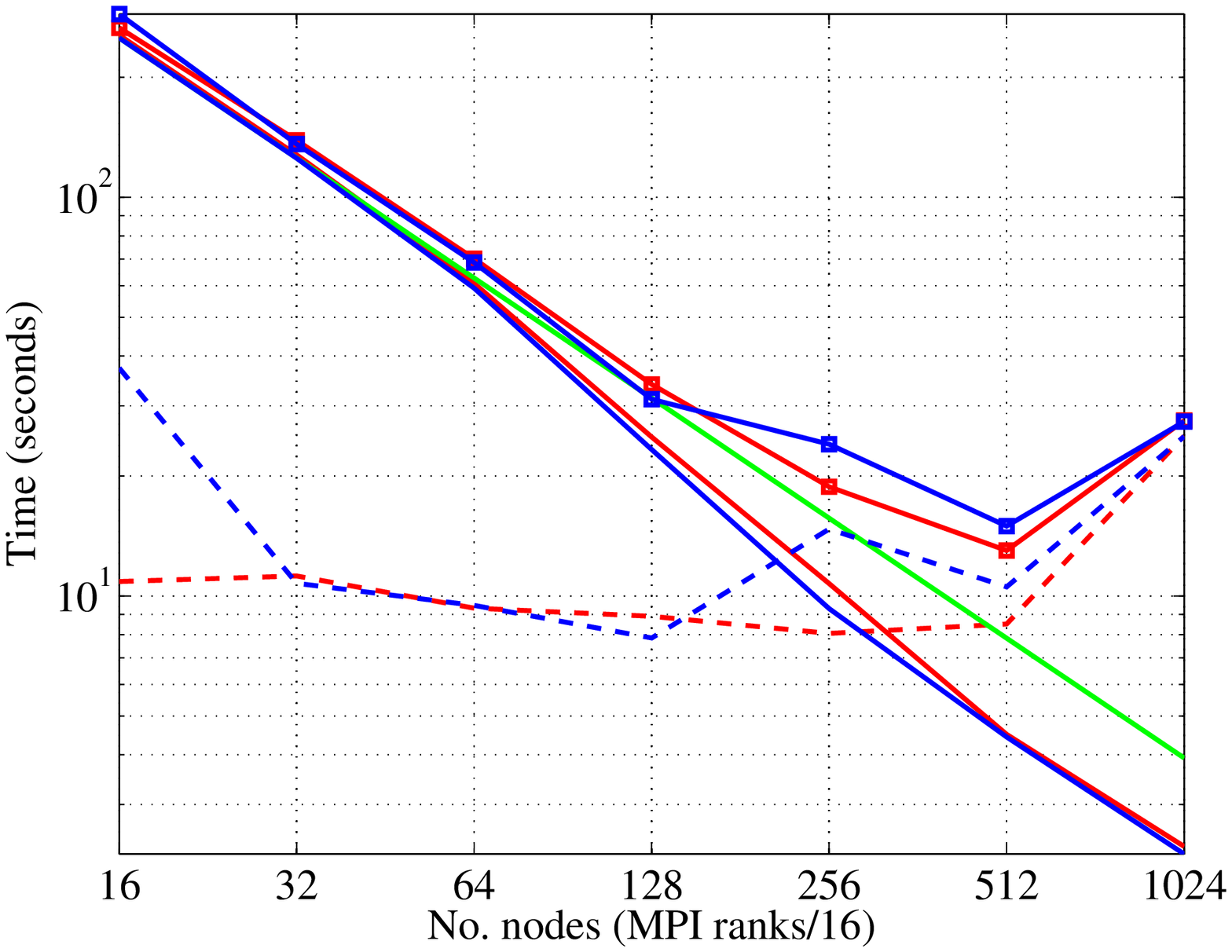}
  }
  \subfigure[$Re_{\tau} = 550$]{
  \includegraphics[width=0.85\linewidth,height=0.2\textheight]{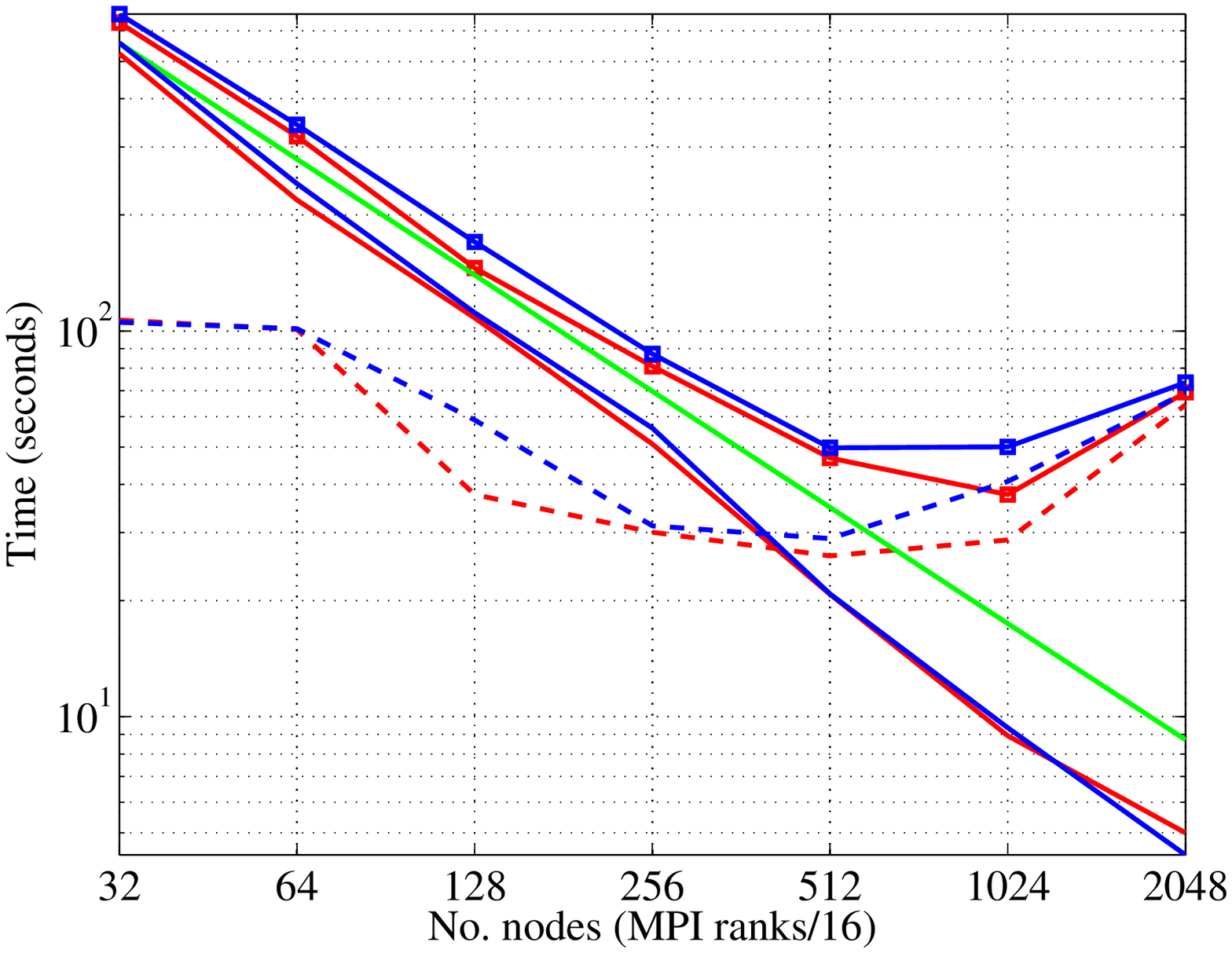}
  }
  \subfigure[$Re_{\tau} = 1000$]{
  \includegraphics[width=0.85\linewidth,height=0.2\textheight]{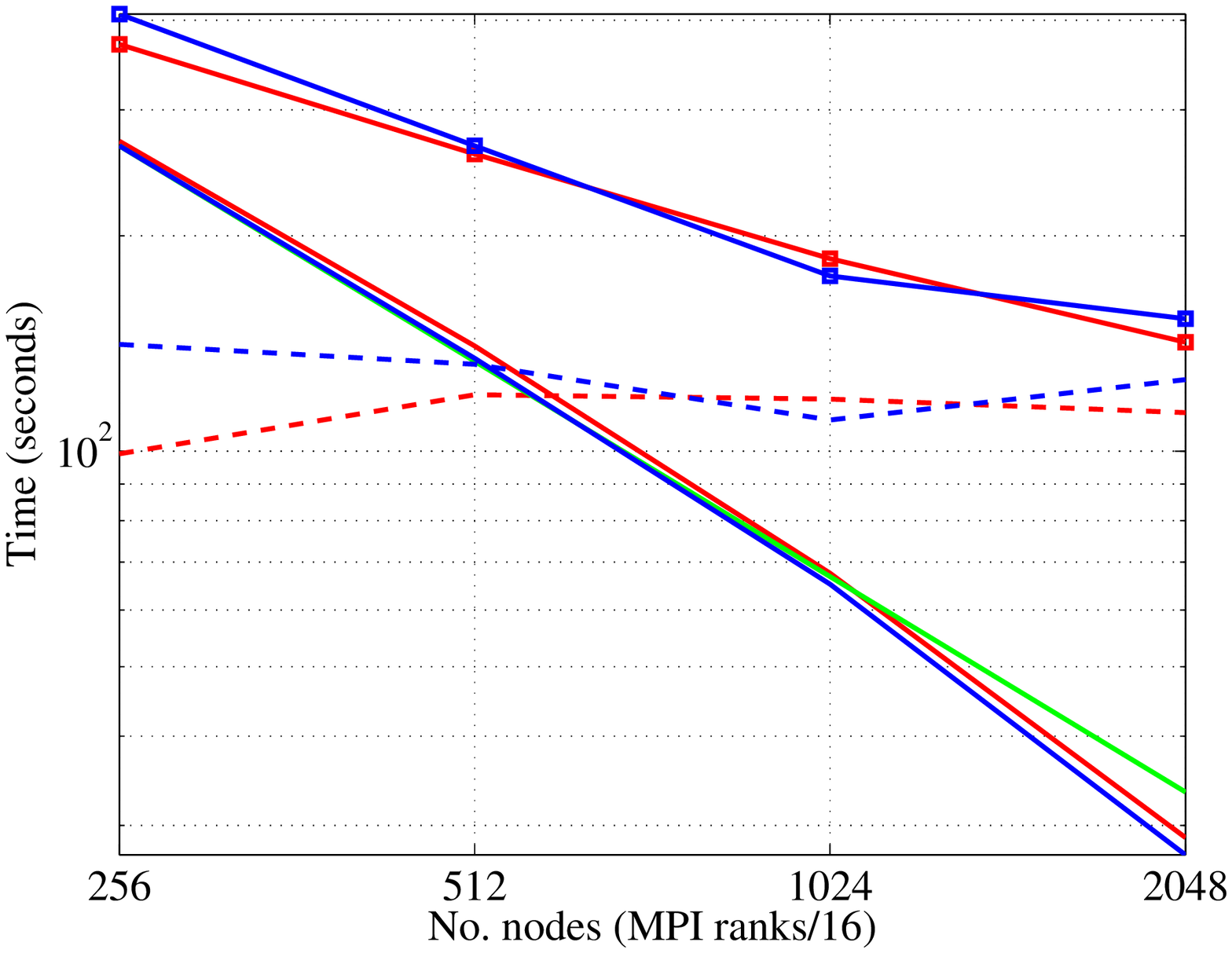}
  }
\caption{Total mean time for 20 timesteps on Titan: AMG ({\color{blue}blue}), XXT ({\color{red}red}), communication (dashed),
  computation (solid),  total time ($\square$), computational linear scaling
  ({\color{green}green}).}
\label{fig:scaling_titan}
\end{figure}

\begin{figure}
  \centering
  \subfigure[$Re_{\tau} = 180$]{
  \includegraphics[width=0.85\linewidth,height=0.2\textheight]{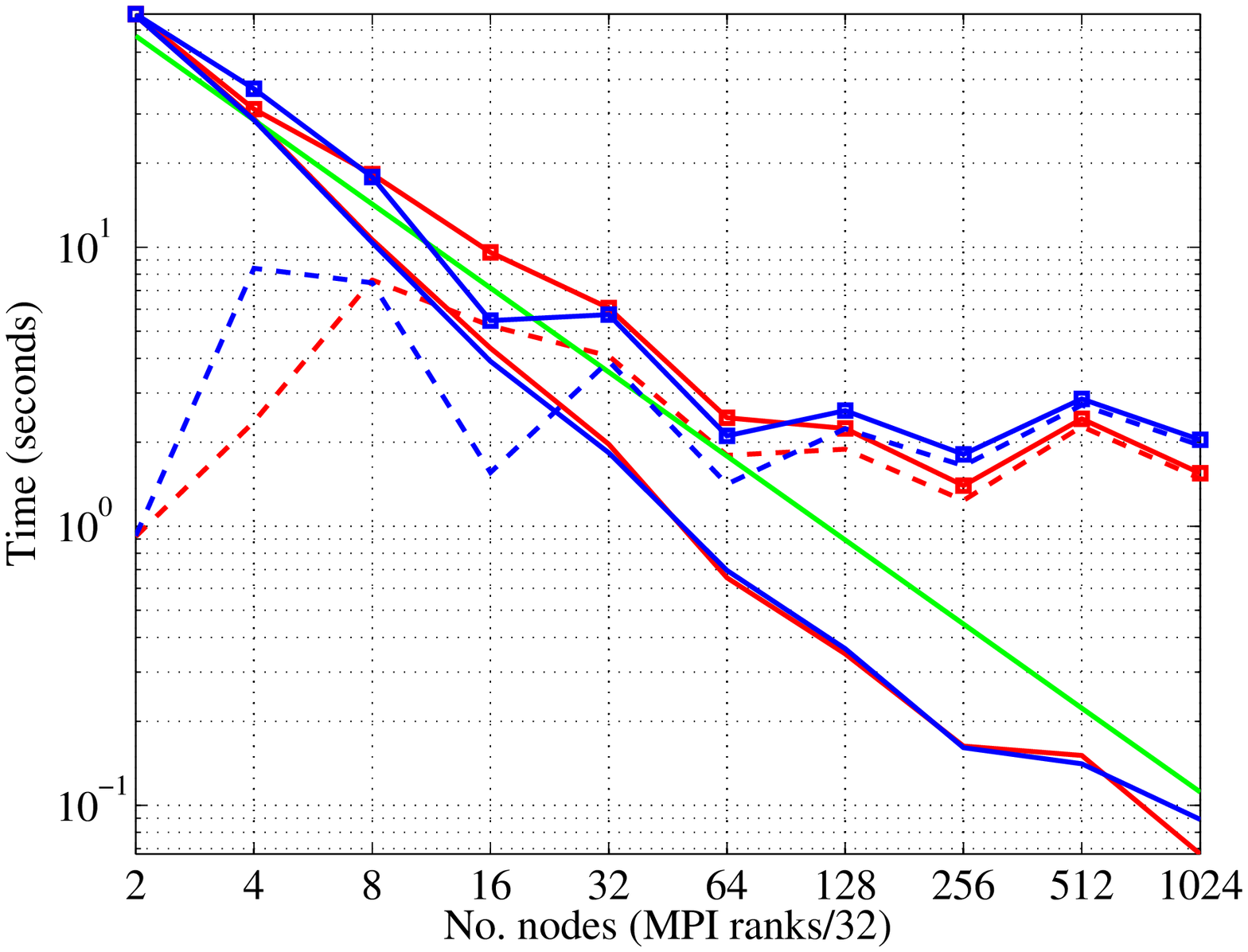}
  }
  \subfigure[$Re_{\tau} = 360$]{
  \includegraphics[width=0.85\linewidth,height=0.2\textheight]{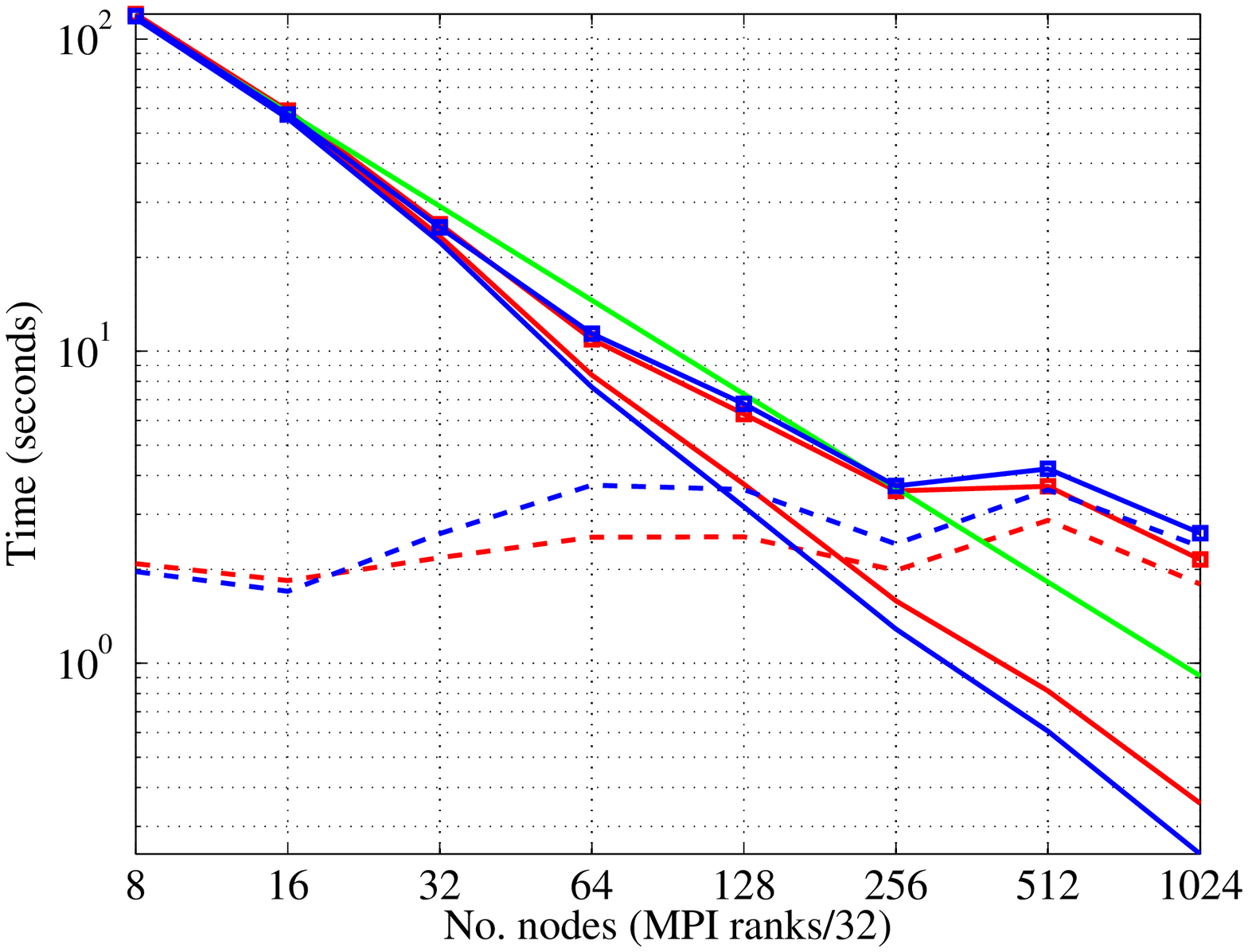}
  }
  \subfigure[$Re_{\tau} = 550$]{
  \includegraphics[width=0.85\linewidth,height=0.2\textheight]{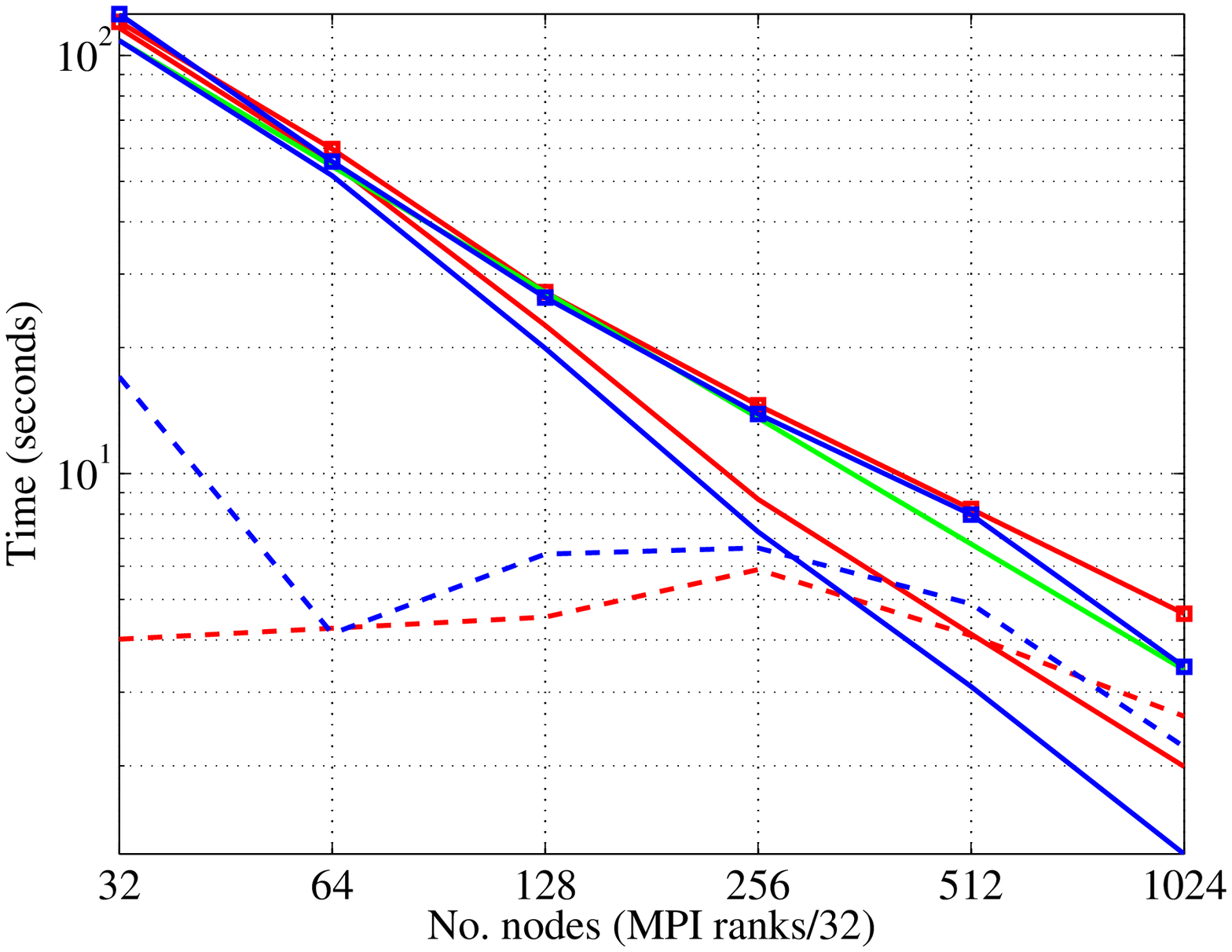}
  }
  \subfigure[$Re_{\tau} = 1000$]{
  \includegraphics[width=0.85\linewidth,height=0.2\textheight]{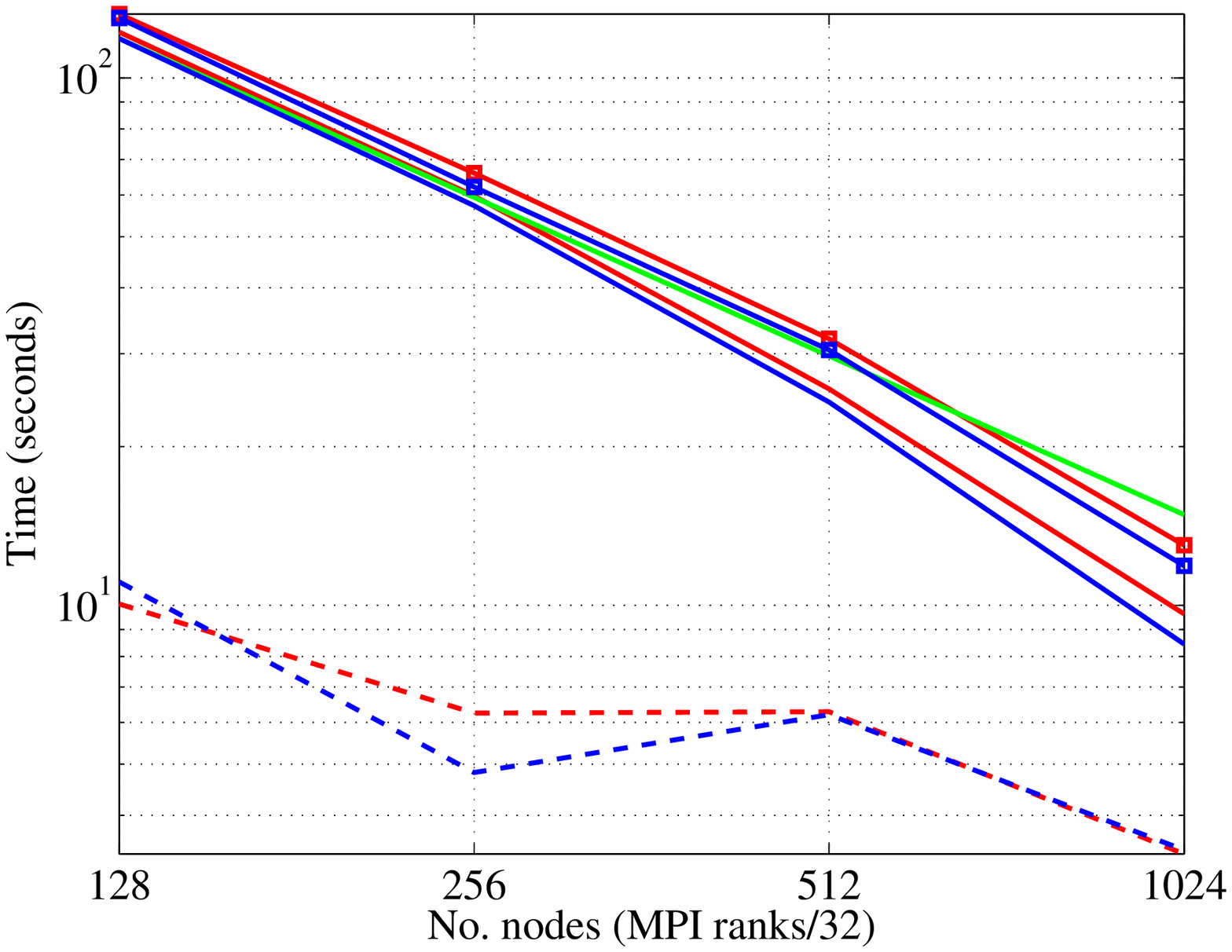}
  }
\caption{Total mean time for 20 timesteps on Beskow: AMG ({\color{blue}blue}), XXT ({\color{red}red}), communication (dashed),
  computation (solid),  total time ($\square$), computational linear scaling
  ({\color{green}green}).}
\label{fig:scaling_beskow}
\end{figure}

The core of the present analysis relies on the data in \reffig{fig:scaling_mira}, \reffig{fig:scaling_titan} and \reffig{fig:scaling_beskow} for each one of the three machines discussed here. 
The compute time and
communication time are illustrated along with the total time for both XXT and AMG across all test cases. 
Since each independent measurement is taken at powers of $2$ number of processes $P$ 
the plots are presented in logarithmic scale. However, in order to help identify linear scaling of the
compute time, the optimal linear scaling line of the compute time was added, i.e.
$$
T_{ideal}(N,P)=\frac{T(N,P_1)P_1}{P}
$$
where $P_1$ is the lowest process count possible for the given problem size. It is noteworthy that we do not compare to linear scaling of the total time as we operate in the strong scaling limit regime, where parallel efficiency is supposed to be well below unity.

\subsection{Strong scaling limit}

For both XXT and AMG the strong scaling limit, i.e. \refeqn{eq:strong}, can be readily extracted from the plots by examining the intersection of the computation time 
and the communication time. For the given intersection point we can identify the values of $\frac{N}{P}$ through linear interpolation and they are reported them in \reftab{tab:stronglimit}.

\begin{table}  
  \caption{Strong Scaling Limit for all 4 test cases on Beskow, Titan and Mira
  in degrees of freedom per core $\frac{N}{P}$ with 2 processes per core on Mira, 1 process per core on Beskow and Titan.}
  \centering
  \begin{tabular}{c||cccc}
    \hline
    {\bf Mira}
    &$Re_{\tau} 180$&$Re_{\tau} 360$&$Re_{\tau} 550$&$Re_{\tau} 1000$\\
    \hline
    XXT&4496&3412&4192&-\\
    AMG&5040&5578&6200&9750\\
    \hline
    \hline
    {\bf Titan}
    &$Re_{\tau} 180$&$Re_{\tau} 360$&$Re_{\tau} 550$&$Re_{\tau} 1000$\\
    \hline
    XXT&9000&24000&65000&228000\\
    AMG&11000&36000&68000&132000\\
    \hline
    \hline
    {\bf Beskow}
    &$Re_{\tau} 180$&$Re_{\tau} 360$&$Re_{\tau} 550$&$Re_{\tau} 1000$\\
    \hline
    XXT&45700&19200&26000& - \\
    AMG&24800&33000&48000& - \\
    \hline
    \hline
  \end{tabular}
  \label{tab:stronglimit}
\end{table}

In practice we observed a strong scaling limit for XXT at roughly $1900<
\frac{N}{P} < 2300$ and for AMG at $2300<\frac{N}{P}<4000$ on Mira, below the $7000$
anticipated in \cite{fischer:scaling}. 

On Beskow, the scalability limit is more difficult to locate with confidence due 
to the high variance in communication times (see \reffig{fig:pingpong_beskow}), 
in particular for small cases. Nevertheless we present it given in 
\reffig{fig:scaling_beskow}, while keeping in mind that this is the result of a 
single run and not averaged across several samples. The scalability limit on 
Beskow is roughly one order of magnitude higher than on Mira in terms of degrees
of freedom per core and is located around $\frac{N}{P} \sim 20,000 - 50,000$. This 
is consistent with the values for the non-dimensional latency $\alpha$ 
from \reftab{tab:alpha_beta}. Indeed Beskow has faster CPUs, thus having a
lower value for $t_a$. This leads to a higher $\alpha$, although the values for
$\alpha*$ are relatively comparable for Titan, Mira and Beskow.

As an intermediate conclusion, we note that the scalability limit on Mira and Beskow is almost independent of the problem size and number of cores that are used. 

On Titan, the scalability limit exhibits a different behavior.
The strong scaling limit for $\frac{N}{P}$ increases
significantly with bigger cases as we see from \reftab{tab:stronglimit}.
The limit goes from $\frac{N}{P} \sim 9000$ to $\frac{N}{P} \sim 228,000$ across the cases $Re_{\tau}=180$ to $Re_{\tau}=1000$ for XXT. Similarly for AMG $\frac{N}{P}  \sim 11,000$ increases sharply to $\frac{N}{P} \sim 132,000$ from the smallest case $Re_{\tau}=180$ to 
$Re_{\tau}=1000$. This limitation cannot be fully explained from the non-dimensional latencies and bandwidths from \reftab{tab:alpha_beta}.

The first plausible explanation is the occasional, random latency spikes seen in the ping-pong tests. If
one process experiences this, the created imbalance has repercussions for all
processes in a parallelized CFD code. We see that these spikes increase communication by
an order of magnitude or higher. The more compute nodes are involved, the higher
the risk of a latency spike. This may partially explain Titan's strong
scaling limit in grid points per process $\frac{N}{P}$ increase with increasing $P$.

Secondly the ping-pong test used to compute the parameters $\alpha$ and $\beta$ is performed on $512$ cores only. During this test, the cores are very likely to lie close to each other on the computer. 
However, this ideal situation does not hold any more when a high number of cores
is considered, as they are probably split on many remote nodes.
A poor interconnect network between cabinets or a high network load on Titan 
could be a valid reason for the observed deterioration and an analysis based on the linear communication model given by  
$\alpha$ and $\beta$  becomes irrelevant on Titan at a high core count.

\subsection{Super-Linearity}

In theory the computational time should match exactly the linear scaling
as work is distributed according to the ratio $\frac{N}{P}$. 

\begin{figure}
  \centering
  \includegraphics[trim=0 4 0 20,clip,width=\linewidth]{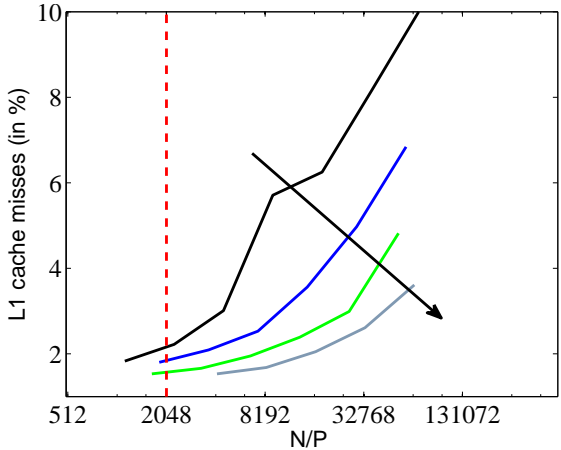}
  \caption{Cache misses on Mira for all 4 test cases (low to high degrees of freedom in direction of the arrow). Percentage of load operations for data and instructions. Dashed red line: L1 cache size. }
   \label{fig:cachemisses}
\end{figure}

All test cases on all the systems show a super-linear scaling. This observation holds true with all timers and profilers switched off.

The usual explanation for super-linear scaling is a sudden decrease of the cache
misses for decreasing $\frac{N}{P}$, as parts of the solver can entirely work on data that lies in the cache. 
To investigate this, we extracted the cache misses on Mira as provided through
HPM by accessing hardware counters (see
\reffig{fig:cachemisses}). Mira is equipped with both a L1 data cache and
instruction cache of 16kB each. The cache misses account for both data and
instruction cache misses. The L2 cache
of 32MB, that is located at the node level, becomes irrelevant as HPM
consistently reported over 97\% cache misses. The L1 cache can be filled with $2048$ double precision
numbers, see
\reffig{fig:cachemisses} the vertical dashed red line which indicates the cache 
size of 16kB below which all gridpoints would theoretically fit into data cache. As we run with 2 processes per core, this would be at roughly $1000$
degrees of freedom per process. Although the data fields achieve those sizes only at the strong scaling
limit we do observe a general decrease in cache misses with decreasing
$\frac{N}{P}$ and thus a better cache exploitation. We cannot explain this behavior as that would require a more
granular analysis of the computational kernels. Measuring the cache misses over
the entire timestep with mixed data and instruction misses, gives us only a
general overview of the cache behavior. In summary, we attribute the super-linear scaling to cache
management and pipelining on the CPU of Mira. We do not possess profiling results 
for Beskow or Titan and cannot study the cache misses there but we assume that the 
reason for super-linearity is similar as for Mira.

\subsection{Comparison between XXT and AMG}

For smaller problem sizes ($Re_{\tau}=180$), XXT slightly outperforms AMG on all machines
for a large number of cores, after the strong scaling limit has been reached. 
For this case, computation time is almost unaffected by the method and the better perfomance
of XXT is attributed to a lower communication. For the cases $Re_{\tau}=360$ and 
$550$ on Mira and Beskow, compute time is noticeably lower for AMG than XXT by
about $5-10\%$. For $Re_{\tau}=550$ on Mira, computation time is also lower
for AMG. This leads to the clear conclusion that AMG for this case on this computer
is systematically better by about $10\%$. However, no such incisive conclusion can be drawn for the other
cases because of varying communication times. For $Re_{\tau}=1000$ on Beskow, 
AMG is once again faster in terms of computation and communication time, but the 
gain is hardly a few percents and we are still far from the strong scaling limit.
On Titan, the difference between AMG and XXT is marginal even if AMG seems overall
slightly slower.

Interesting data are the total number of MPI calls and associated message length
for both methods. In \reftab{tab:xxtamg} we compare XXT and AMG on $Re_{\tau} = 550$ 
at the strong scaling limit of AMG (4096 nodes). The amount of data communicated 
by XXT is by an order of magnitude higher, while AMG uses twice as many MPI calls. 
Therefore, AMG should leap ahead if the systems rely on a low latency network 
combined with a high element count.

\begin{table}
\caption{Number of MPI calls and data communicated on $P=131,072$ at $Re_{\tau}=
550$.}
\centering
\begin{tabular}{l||c|c||c|c}
\hline
&\multicolumn{2} {c||} {\bf AMG}&\multicolumn{2} {c} {\bf XXT}\\
\hline
MPI Routine   &  \#calls  &  bytes  &   \#calls  & bytes \\ 
\hline
MPI\_Isend     &  96638   &       360.5   &  62336     &    6363.8    \\     
MPI\_Irecv     &  96638   &       362.7   &    5916    &    61885.4   \\   
MPI\_Waitall   &  38971   &         0.0   &   56420    &      542.0   \\     
MPI\_Allreduce &  10956   &      5921.0   &   5848     &    10.0   \\     
\hline
Total&252312&&130520\\                                                 
\hline
\end{tabular}
\label{tab:xxtamg}
\end{table}

\subsection{Load balancing}

\begin{figure}
  \centering
  \includegraphics[trim=0 2 0 25,clip,width=\linewidth]{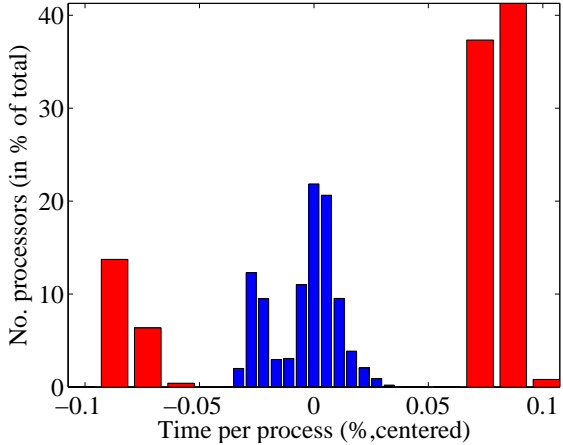}
  \caption{Histogram of the load imbalance at $2.184\times10^9$ degrees of freedom: number of processes per time spent in communication. Red bins: \% of processes at the lowest per core load, blue bins: \% of processes at highest per core load.}
  \label{fig:imbalancehist}
\end{figure}

Beyond the strong scaling limit, the computation time increases and
approaches the linear scaling line again. This is attributed to the load
imbalance as in the extreme case some processes have to work on one element and
some processes on two elements. This can be observed in
\reffig{fig:imbalancehist}
where the imbalance for $Re_{\tau}=1000$ on 32,768 nodes creates two spikes in the
distribution of the execution time. The histogram includes the imbalance of
the workload as well as the resulting imbalance in the communication.

\subsection{Weak scaling}

\begin{figure}
  \centering
  \includegraphics[trim=0 5 0 25,clip,width=\linewidth]{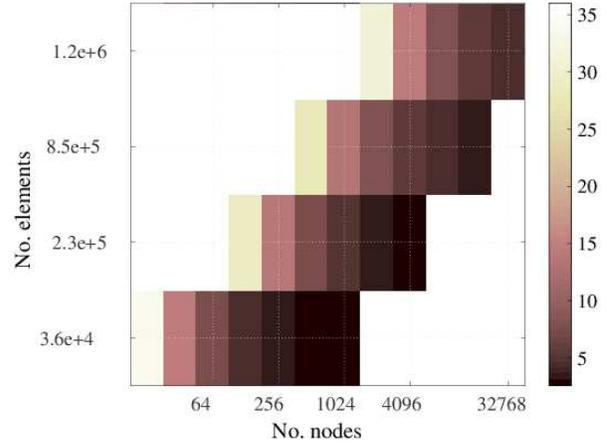}
  \caption{Weak Scaling: For a fixed $\frac{N}{P}$, the same runtime is
  observed across the four test cases, colorbar : total time (seconds).}
  \label{fig:weakscaling}
\end{figure}

As a byproduct of our analysis we can mimic a weak scaling analysis by stacking up 
all problem sizes at every MPI rank count. On Mira the communication is mostly 
latency bound with a small influence from the
bandwidth. This holds also true for peer to peer communication as well as for the
all-reduce (see \reffig{fig:pingpong}). Across the four test cases we can observe 
a weak scaling in \reffig{fig:weakscaling}. It proves that the scaling on Mira is 
mainly dependent on the ratio $\frac{N}{P}$. 

\subsection{Time to solution}
\begin{figure}
  \centering
  \includegraphics[trim=0 2 0 25,clip,width=\linewidth]{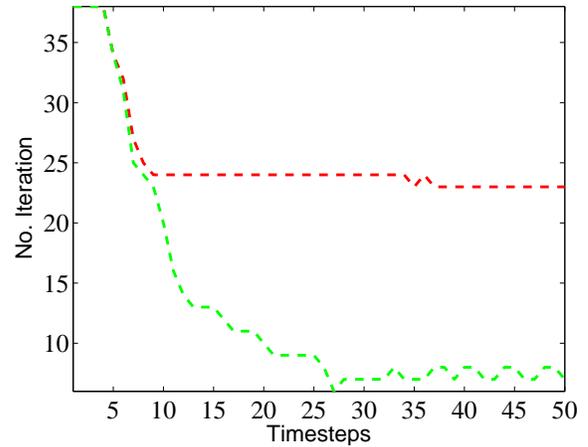}
  \caption{Convergence of solution with increase of projections space $L$ (green: $L=5$, red $L=0$).}
  \label{fig:projections}
\end{figure}

Although we focus on parallelism and performance across a high number of
processes the most important feature of a CFD code remains, in practice, the
ability to minimize the time to solution. The parallelsim is not the sole
contributor to fast time to solution, but also secondary strategies such as
projections mentioned in \refsec{sec:method}. Here we assessed a recent upgraded
implementation of the projections scheme, see \reffig{fig:projections}, where we
note that the number of iterations per solve decreases around 3.5 times as we
increase the projections space $L$. Compared to the reported results in
\cite{Fischer1998} an improved reusability of the projection space data
eliminates the spikes in convergence. As we mentioned before, the memory
footprint of Nek5000 is roughly 500 fields. Thus the additional memory
requirement of the projections is currently of little relevance. 

\section{Conclusion}
Our four test cases from 18 million to 2 billion degrees of freedom were
successfully run on three different petascale system architectures from the
lowest, memory bound, processor count to the granularity limit in order to
assess the strong scaling of Nek5000. We can confirm on Mira that  we can match the theoretical limits established in \cite{fischer:scaling}. However on the Cray systems we observed one to two orders of magnitude
lower strong scaling limits due to higher latency and high noise across the network. 
Our results point at the regimes under which to choose AMG or
XXT; AMG for low latency and high element count, XXT for high latency, high
bandwidth and low element count. The linear communication model proved itself insuficient for explaining the difference in
the strong scaling limit between Titan and Mira. Including a 
quantity for the network noise may improve scaling predictions on noisy systems.
We confirmed that a synchronized and low latency global communication remains crucial for strong scaling of a spectral element based CFD solver.

\section{Acknowledgments}
This research used resources provided by the Swedish National 
Infrastructure for Computing (SNIC) at PDC Centre for High Performance Computing (PDC-HPC).
This research used resources of the Argonne Leadership Computing Facility, which is a DOE Office of Science User Facility supported under Contract DE-AC02-06CH11357.
This research used resources of the Oak Ridge Leadership Computing Facility at 
the Oak Ridge National Laboratory, which is supported by the Office of Science of 
the U.S. Department of Energy under Contract No. DE-AC05-00OR22725. 
This research used resources of the Argonne Leadership Computing Facility, which
is a DOE Office of Science User Facility supported under Contract
DE-AC02-06CH11357.
We thank Scott Parker and Kevin Harms from ALCF at Argonne for their invaluable
suggestions and insights into performance analysis on Mira. We also thank the Nuclear Energy Advanced Modeling and Simulation (NEAMS) program and the Linn{\'e} FLOW Center that funded part of this research.
%
\bibliographystyle{abbrv}
\bibliography{easc2016_offermans}  
%
%

\vskip 40pt
\begin{flushright}
\scriptsize \framebox{\parbox{3.2in}{
The submitted manuscript has been created by the University of Chicago
as Operator of Argonne National Laboratory (``Argonne'') under
Contract No. DE-AC02-06CH11357 with the U.S. Department of Energy.
The U.S. Government retains for itself, and others acting on its
behalf, a paid-up, nonexclusive, irrevocable worldwide license in said
article to reproduce, prepare derivative works, distribute copies to
the public, and perform publicly and display publicly, by or on behalf
of the Government.}} \normalsize
\end{flushright}
\end{document}